\documentclass[]{tMOP2emod}
\def\bra#1{\mathinner{\langle{#1}|}}
\def\ket#1{\mathinner{|{#1}\rangle}}
\def\halfket#1{\mathinner{{#1}\rangle}}

\begin{document}

%\doi{10.1080/0950034YYxxxxxxxx}
% \issn{1362-3044}
%\issnp{0950-0340} \jvol{00} \jnum{00} \jyear{2008} \jmonth{10 January}

\markboth{P.~J.~Windpassinger {\it et al.}}{Squeezing of Atomic Quantum Projection Noise}

%\articletype{GUIDE}

\articletype{Preprint}

\title{Squeezing of Atomic Quantum Projection Noise}

\author{Patrick J. Windpassinger, Daniel Oblak, Ulrich B. Hoff,\\
Anne Louchet, J\"{u}rgen Appel, Niels
Kj{\ae}rgaard$^\ast$\thanks{$^\ast$Corresponding author. Email:
niels.kjaergaard@gmail.com \vspace{6pt}} and Eugene S. Polzik
\\\vspace{6pt}  {\em{Danish National Center for Quantum
Optics---QUANTOP,\\ Niels Bohr Institute, University of Copenhagen,
Denmark}}\\\vspace{6pt}\received{March 24, 2009} }

\maketitle

\begin{abstract}
We provide a framework for understanding recent experiments on
squeezing of a collective atomic pseudo-spin, induced by a homodyne
measurement on off-resonant probe light interrogating the atoms. The
detection of light decimates the atomic state distribution and we
discuss the conditions under which the resulting reduced quantum
fluctuations are metrologically relevant. In particular, we consider
a dual probe scheme which benefits from a cancelation of classical
common mode noise sources such that quantum fluctuations from light
and atoms are the main contributions to the detected signal.
\bigskip

\bigskip

\end{abstract}

\section{Introduction}
During the past year, experiments focusing on pseudo-spin squeezing
via quantum non-demolition (QND) measurements \cite{Kuzmich1998}
have received considerable interest
\cite{Fernholz2008,Teper2008,Takano2009,Schleier-Smith2008,appelarxiv}.
As reported in \cite{appelarxiv}, our group recently demonstrated
3.4
  dB of spectroscopically relevant quantum noise squeezing for up to
  $\sim10^5$ caesium atoms via a QND
  measurement of the population difference between the clock levels. The QND
  measurement was implemented by interferometric detection of the state dependent phase
  shift of an off-resonant dichromatic probe light pulse. In the
  present paper we shall illuminate some of the underlying physics in play
  for such a measurement induced squeezing protocol.
  \section{Atomic projection noise}\label{sec:atomprojnoise}
  \begin{figure}
\begin{center}
\includegraphics[width=\textwidth]{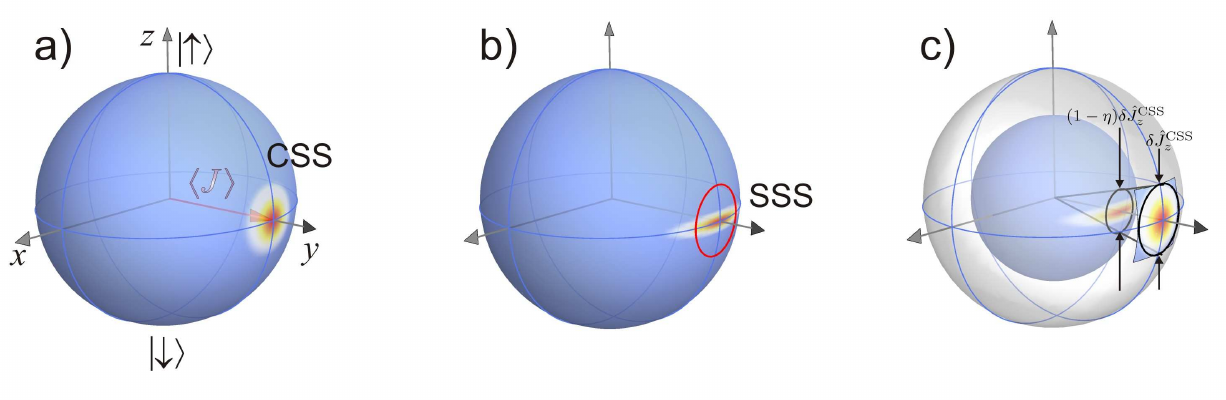}
\caption{\label{figblochspheres} a) Bloch sphere representation of
the coherent spin state (CSS) of Eq.~(\ref{eqprodstate}). The
pointing uncertainty of the Bloch vector with mean $\langle\hat{\bf
J}\rangle$ in pseudo-spin space is illustrated as a shaded disk. b)
Illustration of a spin squeezed state (SSS) with reduced
fluctuations of the $\hat{J}_z$ projection as compared to the CSS.
c) If the protocol for obtaining an SSS from a CSS is accompanied by
decoherence which reduces the radius of the original Bloch sphere by
a factor $(1-\eta)$ the fluctuations of the SSS
$\delta\hat{J}_z^{\rm SSS}$ should be less than
$(1-\eta)\delta\hat{J}_z^{\rm CSS}$ in order for the squeezed
component to increase angular resolution in $yz$-plane
(spectroscopically relevant squeezing). }
\label{sample-figure}
\end{center}
\end{figure}
  Imagine $N$ two-level atoms with energy eigenstates
  $\ket{\downarrow}$ and $\ket{\uparrow}$. With all atoms
  initially in the $\ket{\downarrow}$ state a resonant $\pi/2$-pulse is
  applied on the $\ket{\downarrow}\leftrightarrow\ket{\uparrow}$
  transition preparing each atom in the superposition state
  $(\ket{\downarrow}+\mathrm{i}\ket{\uparrow})/\sqrt{2}$ so that the collective atomic state is described
  by
\begin{equation}\label{eqprodstate}
\ket{\Psi}_{\rm
CSS}=\bigotimes_{k=1}^N(\ket{\downarrow}+\mathrm{i}\ket{\uparrow})_k/\sqrt{2}.
\end{equation}
  A projective measurement of the number of atoms in the
  $\ket{\uparrow}$-state is connected to an expectation value of
  $\langle N_\uparrow\rangle=N/2$ and a variance ${\rm var}(N_\uparrow)=N/4$. It follows that
measurements of the population
  \textit{difference} between $\ket{\downarrow}$ and $\ket{\uparrow}$
  will fluctuate about a zero mean with a variance ${\rm var}(N_\uparrow-N_\downarrow)={\rm var}(2N_\uparrow-N)=4{\rm var}(N_\uparrow)=N$. These
  fluctuations are referred to as atomic projection noise and and they restrict the precision to which we can retrieve
  the value $\pi/2$ for the original angle of rotation from $\ket{\downarrow}^{\bigotimes N}$ based on a measurement of the $\ket{\downarrow}$ and
  $\ket{\uparrow}$ population difference.
Introducing collective atomic operators
\begin{subequations}
\begin{eqnarray}
  \hat{J}_x &=& \frac{1}{2}\sum_{k=1}^N(\ket{\uparrow}\bra{\downarrow}+\ket{\downarrow}\bra{\uparrow})_k, \\
  \hat{J}_y &=& \frac{-\mathrm{i}}{2}\sum_{k=1}^N(\ket{\uparrow}\bra{\downarrow}-\ket{\downarrow}\bra{\uparrow})_k, \\
  \hat{J}_z &=&
  \frac{1}{2}\sum_{k=1}^N(\ket{\uparrow}\bra{\uparrow}-\ket{\downarrow}\bra{\downarrow})_k,
\end{eqnarray}
\end{subequations}
which may be shown to obey angular momentum commutation relations
$[\hat{J}_k,\hat{J}_l]=i\epsilon_{klm}\hat{J}_m$, we can rewrite
$\ket{\Psi}_{\rm CSS}$ in terms of simultaneous eigenstates of
$\hat{J}_z$ and $\hat{\bf J}^2$
\begin{equation}\label{eqdickeexp}
\ket{\Psi}_{\rm
CSS}=\frac{1}{2^{N/2}}\sum_{M=-N/2}^{N/2}\left(\begin{array}{c}
                                                                      N \\
                                                                      N/2+M
                                                                    \end{array}
\right)^{1/2}\ket{N/2,M},
\end{equation}
i.e., the (sub-)set of Dicke states \cite{Dicke1954,Mandel1997}
fulfilling $\hat{J}_z\ket{N/2,M}=M\ket{N/2,M}$ and $\hat{\bf
J}^2\ket{N/2,M}=(N/2)(N/2+1)\ket{N/2,M}$. $\ket{\Psi}_{\rm CSS}$ can
be obtained by a rotation $(\theta=\pi/2,\varphi=0)$ of the Dicke
state $\ket{N/2,-N/2}$ in angular momentum space and is referred to
as a Bloch state, an atomic coherent state, or a coherent spin state
(CSS). The polar angle $\theta$ and azimuth $\varphi$ parameterize
the Bloch sphere of radius $N/2$. As illustrated in
Fig.~\ref{figblochspheres}(a), we can represent the
  uncertainty associated with a projective measurement on $\ket{\Psi}_{\rm CSS}$ as a disk of radius $\sqrt{N}$ on
  the Bloch sphere, where the uncertainty disk is centered on the tip of the Bloch vector $\langle\hat{\bf J}\rangle$. We conclude that Ramsey
  spectroscopy which
  essentially relies on determining such angular rotations of a
  Bloch vector is ultimately limited by atomic projection noise \cite{Santarelli1999}. The
  prospect of surpassing this so-called standard quantum limit (SQL)
  motivates work towards the production of squeezed atomic
  states [see Fig.~\ref{figblochspheres}(b)] displaying a reduction of quantum projection noise in a given
  measurement basis \cite{WINELAND1992}.

\section{Projection noise measurements and squeezing}
In order to infer squeezing of atomic quantum noise it is obviously
a requirement to have a measurement scheme with sufficient
sensitivity to reveal the projection noise limit for a CSS in the
first instance. However, if we have the ability to perform such a
projection noise limited measurement on a CSS and, moreover, the
measurement has a QND character we have already succeeded in
producing a squeezed atomic state. The very act of measuring defines
one component of the collective pseudo-spin (say, the population
difference) to a precision better than the SQL for a CSS. Such a QND
measurement can be implemented by determining the refractive index
of the atomic ensemble using off-resonant probe light
\cite{Kuzmich1998}.

\subsection{Refractive index measurements via phase shift of light}
\begin{figure}
\begin{center}
\includegraphics[width=5cm]{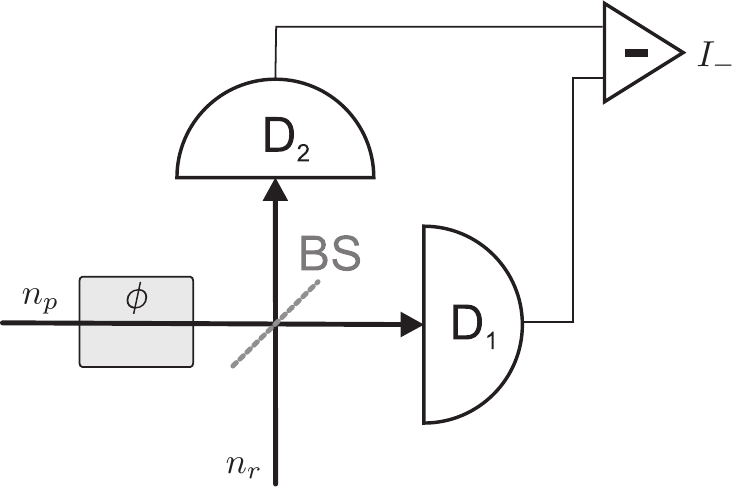}
\caption{\label{fig2} Balanced homodyne measurement of the phase shift $\phi$ between a probe and reference light field as the difference signal $I_-$ between photo detectors D1 and D2.}%
\label{sample-figure}
\end{center}
\end{figure}
When a probe beam of light of wavelength $\lambda$ propagates a
distance $l$ through a medium of refractive index $n$ it acquires a
phase shift $\phi=\frac{2\pi l}{\lambda}(n-1)$. The phase shift can
be measured in a balanced homodyne detection setup (see
Fig.~\ref{fig2}) where in essence the phase of the probe beam is
compared to the phase of a reference beam (local oscillator).
Describing the probe and reference pulses of light by coherent
states with average photon numbers $n_p$ and $n_r$, respectively,
the differential signal from the two detectors yields (on average) a
signal (see appendix \ref{app:phase})
\begin{equation}\label{eqdiffcurrent}
\langle I_-\rangle= 2\sqrt{n_rn_p}\sin\phi.%\equiv2\sqrt{n_rn_p}\phi
\end{equation}
with a second moment given by
\begin{equation}\label{eqdiffcurrentvariance}
\langle{I}_-^2\rangle= n_r+n_p+4n_rn_p\sin^2\phi.
\end{equation}
For a \textit{constant} phase shift the differential detector signal
fluctuates as ${\rm
var}\left(I_-\right)=\langle{I}_-^2\rangle-\langle{I}_-\rangle^2=n_r+n_p$
independent of $\phi$ which is nothing but the shot noise of light.
For a fluctuating phase $\phi$ we obtain using the law of total
variance
\begin{eqnarray}\label{eqvardiffcurrent}
  {\rm var}\left(I_-\right) &=& \left\langle {\rm var}\left(I_-|\phi\right)\right\rangle+{\rm var}\left(\langle
  I_-|\phi\rangle\right)\nonumber\\&=& n_r+n_p+{\rm
  var}\left(2\sqrt{n_rn_p}\sin\phi\right)\approx n_r+n_p+{\rm
  var}\left(2\sqrt{n_rn_p}\phi\right)\nonumber\\&=&n_r+n_p+4n_rn_p{\rm
  var}(\phi),
\end{eqnarray}
where we have assumed that $\phi\ll \pi/2$ so that $\sin
\phi\approx\phi$.

\subsubsection{Projection noise measurements on a CSS using one and two probes}

Suppose now that the refractive medium is made up by $N$ atoms in
a CSS as described in section \ref{sec:atomprojnoise} and moreover
that the probe laser phase shift is proportional to the number
$N_\uparrow$ of atoms in the $\ket{\uparrow}$-state:
\begin{equation}\label{eqphase}
    \phi_\uparrow=k_\uparrow N_\uparrow+\phi_0,
\end{equation}
where $\phi_0$ is a fluctuating background phase not related to the
atoms. By off-setting the phase of the reference beam by $\langle
\phi_0\rangle+\langle k_\uparrow\rangle N/2$, the
differential signal $I_-$ (representing the phase difference between
the reference and probe fields) fluctuates about zero and the approximation $\sin\phi\approx\phi$ is valid.
Equation~(\ref{eqvardiffcurrent}) then yields (cf.
appendix~\ref{app:noiseone})
\begin{equation}\label{eqsingleprobe}
  {\rm var}\left(I_-^{(\uparrow)}\right) = n_r+n_p+n_rn_p\left\{\left[\langle k_\uparrow\rangle^2+{\rm var}\left(k_\uparrow\right)N+{\rm var}\left(k_\uparrow\right)\right]N+4{
  \rm
  var}\left(\phi_0\right)\right\}.
\end{equation}

We next introduce an additional probe laser with mean photon number $m_p$
sensitive to the population $N_\downarrow$ of the $\ket{\downarrow}$
state. This probe is homodyned with a reference (mean photon number
$m_r$) having a local oscillator phase such that
\begin{eqnarray}
\phi_\downarrow=-(k_\downarrow N_\downarrow+\phi_0).
\end{eqnarray}
For $k_\downarrow=k_\uparrow\equiv k$ and $n_rn_p=m_rm_p$ we get
(cf. appendix~\ref{app:noisetwo})
\begin{eqnarray}\label{eqsnpluspn}
  {\rm var}\left(I_-^{(\uparrow\downarrow)}\right) &=& n_r+n_p+m_r+m_p+4n_rn_p\left[\langle k\rangle^2+{\rm
  var}(k)\right]N\nonumber\\
  &\approx&\underbrace{n_r+n_p+m_r+m_p}_{\text{shot noise}}+\underbrace{4n_rn_p\langle k\rangle^2N}_{\text{projection noise}},
\end{eqnarray}
where the approximation in the second line is valid if ${\rm var
}(k)\ll \langle k\rangle^2$. The variance of the differential
detector signal for dual probe interrogation Eq.~(\ref{eqsnpluspn})
displays a shot noise term and a projection noise term, the latter
scaling linearly with the atom number $N$. In contrast to single
probe interrogation Eq.~(\ref{eqsingleprobe}) there is a cancelation
of background phase fluctuations not relating to atoms as well as
the component quadratic in the number of atoms.

\subsubsection{Two-color dispersive probing
scheme}\label{sectwocolor} 
\begin{figure}
\begin{center}
\includegraphics{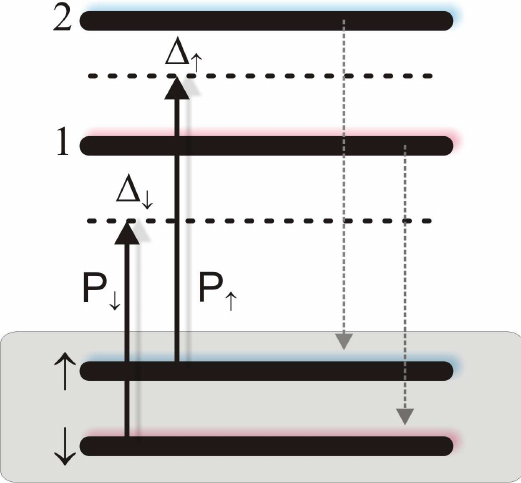}
\caption{Level diagram for a two color dispersive probing scheme. The populations of the
levels $\downarrow$ and $\uparrow$ are measured with optical probes which interacts off-resonantly
with the auxiliary levels 1 and 2, respectively.\label{figleveldiag}}%
\end{center}
\end{figure}

In practice, to achieve a dual probe
interrogation scheme \cite{Saffman2009} we can consider a situation
as shown in Fig. \ref{figleveldiag}. Here the frequency of the probe
$P_\uparrow(P_\downarrow)$ is detuned
$\Delta_\uparrow(\Delta_\downarrow)$ from resonance of the closed
optical transition between the $\ket{\uparrow}(\ket{\downarrow})$
state and an auxiliary quantum state $\ket{1}(\ket{2})$.
If the detunings are large compared to the respective line widths
$\Gamma_\uparrow(\Gamma_\downarrow)$ of these two transitions the
dispersive interaction responsible for the phase shift of light from
atoms will have a simple $1/\Delta_{\uparrow(\downarrow)}$
dependence, i.e.:
\begin{eqnarray}\label{eqcouplingcst}
  k_\downarrow = \frac{c_{\downarrow,1}}{\Delta_\downarrow}\quad ,\quad
  k_\uparrow = \frac{c_{\uparrow,2}}{\Delta_\uparrow},
\end{eqnarray}
where $c_{\downarrow,1}$ and $c_{\uparrow,2}$ are constants
comprising atomic parameters and the probing geometry. For
comparable detunings $\Delta_\uparrow\approx\Delta_\downarrow$ the
task of keeping $k_\uparrow=k_\downarrow$ translates into
stabilizing the difference in frequency between the two probe fields
which is experimentally feasible \cite{appelphaselock} when the
transition $\ket{\downarrow}\leftrightarrow\ket{\uparrow}$ is in the
radio frequency domain (which is the case for hyperfine split
states). The requirement ${\rm var }(k)\ll \langle k\rangle^2$ is
then fulfilled if the absolute frequency fluctuations of
$P_\uparrow$ is $\ll \Delta_\uparrow$ which is also comfortably met
for $\Delta\gtrsim 100$~MHz, e.g. by referencing $P_\uparrow$ to an
atomic transition. In contrast, single probe interrogation poses a
criterion ${\rm var }(k)\ll \langle k\rangle^2/N$ which, e.g.,
requires the probe frequency fluctuations to be $\ll
\Delta_\uparrow/\sqrt{N}$. This quickly becomes a serious challenge
for ensembles containing hundred thousands of atoms.

\subsubsection{Measurement based quantum noise reduction}
When no atoms are present the differential detector signal will
fluctuate about zero with a variance given by the shot noise term in
Eq.~(\ref{eqsnpluspn}) $n_{\rm sn}\equiv n_r+n_p+m_r+m_p$. If we
introduce atoms in a Dicke state $\ket{N/2,M}$, i.e. a
\textit{fixed} atomic phase shift, the $I_-$ reading would continue
to fluctuate with a variance $n_{\rm sn}$, but now about a mean
value $4\sqrt{n_rn_p}kM\equiv gM$. The probability distribution of
the photonic signal $I_-$ conditioned on the atomic state
$\ket{N/2,M}$ is then
\begin{equation}\label{eqprobdist}
P\left(I_-=n\big|\ket{N/2,M}\right)=\frac{1}{\sqrt{2\pi n_{\rm
sn}}}\exp\left[-\frac{(n-gM)^2}{2n_{\rm sn}}\right],
\end{equation}
where the Skellam distributed differential photon current has been
approximated by a Gaussian. With the atoms prepared in the CSS
Eq.~(\ref{eqdickeexp}) which is a distribution over Dicke states,
the probability for a given Dicke state is
\begin{equation}\label{eqdickeprob}
P(\ket{N/2,M})=\left | \bra{N/2,M}\halfket{\Psi}_{\rm CSS}\right
|^2=\frac{1}{2^N}\left(\begin{array}{c}
                                                                      N \\
                                                                      N/2+M
                                                                    \end{array}
\right).
\end{equation}
Using Bayes' rule the distribution over $M$-states given a detection
event $I_-=n$ can be inferred
\begin{eqnarray}\label{eqproduct}
  P\left(\ket{N/2,M}\big|I_-=n\right)&\propto&  \exp\left[-\frac{(n-gM)^2}{2n_{\rm
    sn}}\right]\exp\left[-\frac{M^2}{N/2}\right]\nonumber\\&\propto&\exp\left[-\frac{\left(M-\frac{gN/4n_{\rm sn}}{1+g^2N/4n_{\rm sn}}n\right)^2}{\frac{N/2}{1+g^2N/4n_{\rm
    sn}}}\right].
\end{eqnarray}
Compared to the initial CSS $\ket{\Psi}_{\rm CSS}$ this $\hat{J}_z$
projection distribution has a variance reduced by a factor
$1+g^2N/4n_{\rm sn}\equiv1+\kappa^2$ and is centered at the value
$n\kappa^2/g(1+\kappa^2)$. We refer to this as a conditionally
squeezed state since based on the measurement outcome $I_-=n$ we can
predict the outcome of a subsequent $\hat{J}_z$ measurement to a
precision better than the SQL for a CSS.
\begin{figure}
\begin{center}
\includegraphics[width=\textwidth]{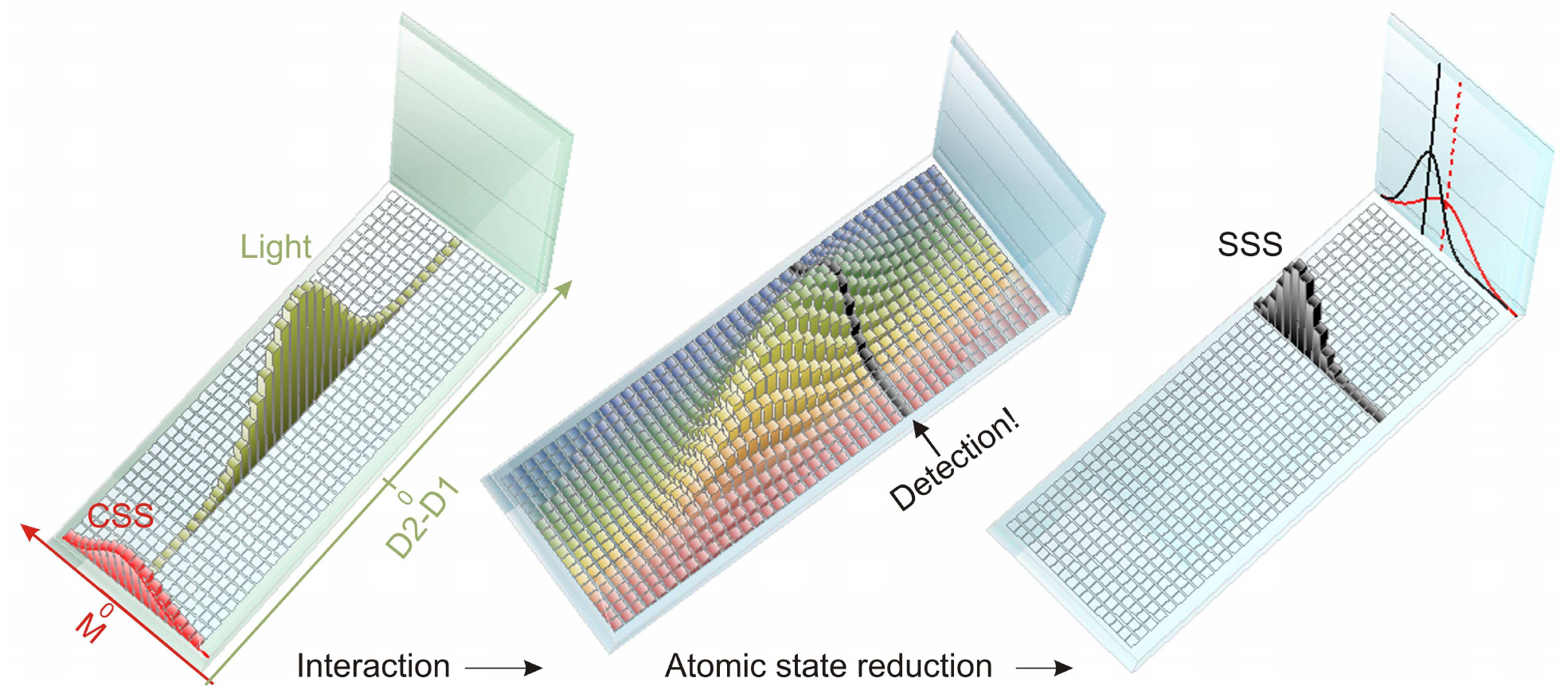}
\caption{Graphical illustration of atomic projection noise reduction
based upon the state reduction which happens after the detection of
phase shifted probe light. The interaction between light and a CSS
(left pane) gives rise to the joint distribution in the center pane.
Upon detection of a differential photo detector signal $I_-=D2-D1$
the atomic distribution over $M$-states is decimated as illustrated
in the right pane. This state has $\hat{J}_z$-projection which is
defined beyond the SQL as given by the initial CSS.
\label{figfancy}}%

\end{center}
\end{figure}
In Fig.~\ref{figfancy} we illustrate the process of quantum noise
reduction (and the shift of the projection distribution) as a result
of a particular measurement outcome of $I_-$. The parameter
$\kappa^2\equiv g^2N/4n_{\rm sn}=4n_rn_pk^2N/(n_r+n_p+m_r+m_p)$
characterizes the amount of squeezing that is achieved. From
Eq.~(\ref{eqsnpluspn}) we see that this is nothing but the ratio of
atomic projection noise to the shot noise of light encountered in
the signal $I_-$. In the case of strong local oscillators of equal intensities $n_r=m_r\gg n_p=m_p$, we have
\begin{equation}\label{eqkappa}
\kappa^2=2k^2Nn_p.
\end{equation}

\subsubsection{Decoherence from spontaneous scattering}
\begin{figure}
\begin{center}
\includegraphics[width=\textwidth]{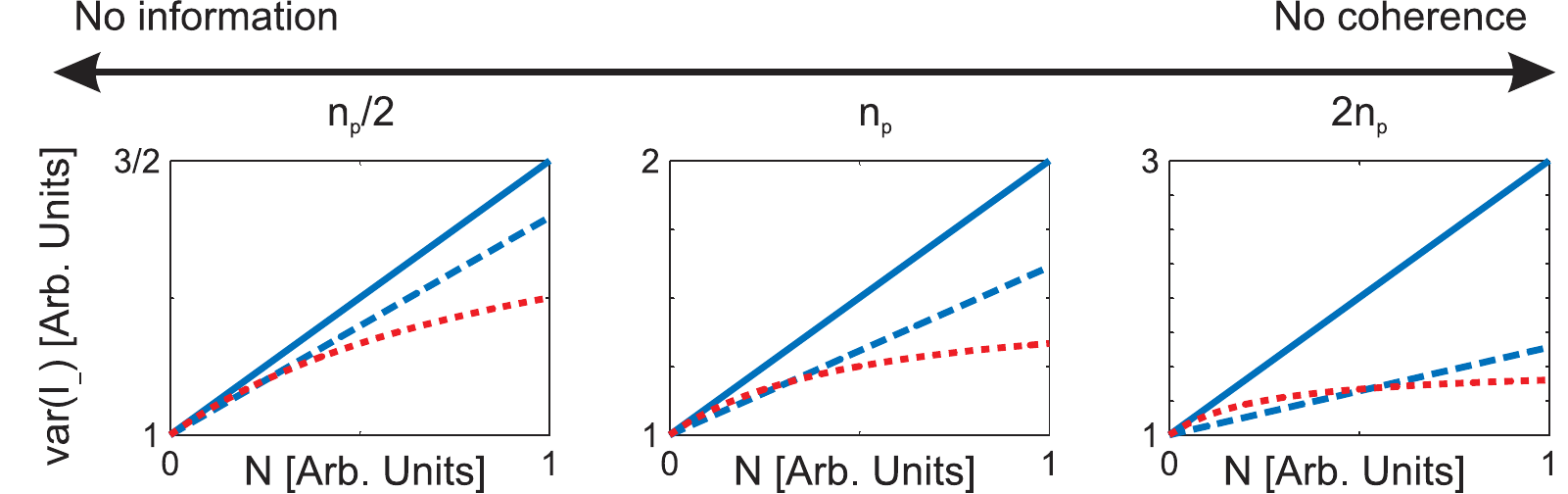}
\caption{Squeezing and the effect of varying the number of probe photons. The full lines show
the projection noise (variance) for a measurement on a CSS which scales linearly with
atom number $N$.  The outcome of a subsequent measurement can be
predicted within the variance given by the dotted line. This
conditionally reduced variance must be below the projection noise
benchmark \textit{scaled down} by a factor $(1-\eta)^2$ (dashed
line) for the noise squeezing to be metrologically relevant [cf.
Fig.~\ref{figblochspheres}(c)].\label{fignoisegraphs}}%
\end{center}
\end{figure}

According to Eq.~(\ref{eqkappa}) a doubling of the probe photon
number doubles the $\kappa^2$ parameter. Hence one might expect
increased squeezing and an improved angular definition of the the
Bloch vector. However, this is not generally true since the
dispersive coupling as characterized by Eq.~(\ref{eqcouplingcst}) is
inevitably accompanied by spontaneous photon scattering with a
direct link provided by the Kramers-Kronig relations. If a fraction
$\eta$ of coherent atoms spontaneously scatters a photon, the length
of the Bloch vector after light-atom interaction is
$(1-\eta)|\langle\hat{\bf J}\rangle|$. Hence, for an initial CSS
with fluctuations $\delta\hat{J}_z^{\rm CSS}$, the fluctuations of
the resulting squeezed state needs to be less than
$(1-\eta)\delta\hat{J}_z^{\rm CSS}$ in order to decrease the
\textit{angular} uncertainty of the Bloch vector in $yz$-plane [See
Fig. \ref{figblochspheres}c]. Thus, the variance of the squeezed
state must be less than $(1-\eta)^2\delta^2\hat{J}_z^{\rm CSS}$ for
\textit{metrologically relevant} squeezing, i.e,
$(1+\kappa^2)^{-1}<(1-\eta)^2$. With $\alpha=(\Gamma k/\Delta)(N/2)$
being the absorption coefficient for each probe and neglecting
depletion, a total of $2n_p\alpha=n_p(\Gamma k/\Delta)N$ photons are
spontaneously scattered and

\begin{equation}\label{eqeta}
 \eta=\frac{\Gamma  k}{\Delta} n_p.
 \end{equation}
 The tradeoff between information gained via Eq.~(\ref{eqkappa}) and
 coherence lost via Eq.~(\ref{eqeta}) in
 terms of number of probe photons is illustrated in Fig.~\ref{fignoisegraphs}. This leads to an optimal decoherence parameter $\eta$ \cite{Saffman2009}. We note
 that since (via $k$) \textit{both} $\kappa^2$ and $\eta$ are $\propto
 1/\Delta^2$, the detuning enters the squeezing optimization
 completely equivalently to $n_p$. So despite the fact that the absorption falls off as $1/\Delta^2$ while
 the index of refraction only falls off as $1/\Delta$ the maximally
 achievable metrologically relevant squeezing for a given number of atoms
 $N$ does not increase with the probe laser detuning: in principle, the maximum can be obtained for any
 detuning by adjusting the probe photon number accordingly. In practical experiments, to reduce the effect of probe frequency
 fluctuations a certain detuning may, however, be required (cf. section
 \ref{sectwocolor}).
\section{Conclusion}
In conclusion, we have elaborated on some important perspectives of
our recent experiments \cite{appelarxiv} demonstrating
spectroscopically relevant quantum noise squeezing on the Cs clock
transition. Here a conditionally squeezed state was produced by
performing a QND measurement on the atomic pseudo-spin projection
using a two-color dispersive probing scheme. The two-color protocol
for which we have evaluated the noise terms benefits from several
levels of common-mode rejection. We have provided a description of
the atomic quantum state reduction resulting from the measurement on
off-resonant probe light after QND interaction with a CSS. This
generally gives rise to a state with reduced fluctuations of an
atomic pseudo-spin component as compared to the initial CSS. We have
discussed the conditions for this noise reduction to be of
metrological relevance and aspects of the tradeoff between atomic
decoherence and information gain.
\section*{Acknowledgements}
This work was funded by the Danish National Research Foundation and
the EU grants COMPAS, EMALI, and QAP. N.K. acknowledges support from
the Danish Natural Science Research Council through a Steno
Fellowship.
 \label{lastpage}
%\bibliographystyle{tMOP2}
%\bibliography{C:/rabibibab}

\appendices
\section{Phase detection}\label{app:phase}
Referring to Fig.~\ref{fig2} and assuming a 50/50 beam splitter the
differential detector output is

\begin{eqnarray}\label{eqhomo}
\hat{I}_-&=&\left [ \frac{1}{\sqrt{2}}(\hat{a}_p+\mathrm{i}\hat{a}_r )\right
]^\dagger\left [ \frac{1}{\sqrt{2}}(\hat{a}_p+\mathrm{i}\hat{a}_r )\right ]
-\left [ \frac{1}{\sqrt{2}}(\mathrm{i}\hat{a}_p+\hat{a}_r )\right
]^\dagger\left [ \frac{1}{\sqrt{2}}(\mathrm{i}\hat{a}_p+\hat{a}_r )\right
]\nonumber\\
&=&\mathrm{i}(\hat{a}_p^\dagger\hat{a}_r-\hat{a}_r^\dagger\hat{a}_p),
\end{eqnarray}
where $\hat{a}_p$ and $\hat{a}_r$ denote the annihilation operators
for the probe and reference field, respectively. The reference and
probe fields are described by coherent states of amplitudes
$\alpha_r$ and $\alpha_p$ with mean photon numbers
$n_r=\langle\hat{a}_r^\dagger\hat{a}_r\rangle=|\alpha_r|^2$ and
$n_p=\langle\hat{a}_p^\dagger\hat{a}_p\rangle=|\alpha_p|^2$,
respectively. Taking, without loss of generality, the coherent state
amplitude for the reference field to be real
$\alpha_r=\alpha_r^\ast=\sqrt{n_r}$ we obtain
\begin{eqnarray}
 \langle\hat{I}_- \rangle
 &=&\langle\alpha_r,\alpha_p|\mathrm{i}(\hat{a}_p^\dagger\hat{a}_r-\hat{a}_r^\dagger\hat{a}_p)|\alpha_r,\alpha_p\rangle=\mathrm{i}(\alpha_p^\ast\alpha_r-\alpha_r^\ast\alpha_p)\nonumber\\&=&2\sqrt{n_r}\left(\frac{\alpha_p-\alpha_p^\ast}{2\mathrm{i}}\right
 )=2\sqrt{n_r}\sqrt{n_p}\sin\phi,
\end{eqnarray}
where $\phi$ is the phase of the probe field with respect to the the
reference field. Calculating the second moment of $\hat{I}_-$ we get
\begin{eqnarray}
 \langle\hat{I}_-^2 \rangle
 &=&\langle\alpha_r,\alpha_p|\left
 [\mathrm{i}(\hat{a}_p^\dagger\hat{a}_r-\hat{a}_r^\dagger\hat{a}_p)\right]^2|\alpha_r,\alpha_p\rangle\nonumber\\&=&\langle\alpha_r,\alpha_p|\hat{a}_p\hat{a}_p^\dagger\hat{a}_r^\dagger\hat{a}_r+\hat{a}_p^\dagger\hat{a}_p\hat{a}_r\hat{a}_r^\dagger-\hat{a}_p^\dagger\hat{a}_p^\dagger\hat{a}_r\hat{a}_r-\hat{a}_p\hat{a}_p\hat{a}_r^\dagger\hat{a}_r^\dagger|\alpha_r,\alpha_p\rangle\nonumber\\&=&(1+n_p)n_r+n_p(1+n_r)-2n_r\frac{{\alpha_p^\ast}^2+\alpha_p^2}{2}
 \nonumber\\&=&n_r+n_p+2n_rn_p(1-\cos 2\phi)=n_r+n_p+4n_rn_p\sin^2
 \phi.
\end{eqnarray}
\section{Noise terms}\label{app:noise}
In order to derive Eqns.~(\ref{eqsingleprobe}) and
(\ref{eqsnpluspn}) from Eq.~(\ref{eqvardiffcurrent}) we must
calculate ${\rm var}(\phi_\uparrow)$ and ${\rm
var}(\phi_\uparrow+\phi_\downarrow)$, respectively.
\subsection{Single color probe}\label{app:noiseone}
Assuming $k_\uparrow$,$N_\uparrow$, $\phi_0$ to be independent we
obtain
\begin{equation}\label{eqsingleprobeapp}
  {\rm var}\left(\phi_\uparrow\right) ={\rm var}\left(k_\uparrow N_\uparrow+\phi_0\right)={\rm var}\left(k_\uparrow N_\uparrow\right)+{\rm
  var}\left(\phi_0\right),
\end{equation}
\begin{eqnarray}\label{eqsingleprobeapp2}
  {\rm var}\left(k_\uparrow N_\uparrow\right) &=&\langle k_\uparrow\rangle^2\underbrace{{\rm var}\left(N_\uparrow\right)}_{N/4}+{\rm var}\left(k_\uparrow\right) \underbrace{\langle N_\uparrow\rangle^2}_{N^2/4}+{\rm var}\left(k_\uparrow\right) \underbrace{{\rm
  var}\left(N_\uparrow\right)}_{N/4}\nonumber\\&=&\left[\langle k_\uparrow\rangle^2+{\rm var}\left(k_\uparrow\right) N+{\rm
  var}\left(k_\uparrow\right)\right]N/4,
\end{eqnarray}
where use of the CSS properties ${\rm var}(N_\uparrow)=N/4$ and
$\langle N_\uparrow\rangle^2=N^2/4$ has been made.

\subsection{Two color probe}\label{app:noisetwo} For $k\equiv k_\uparrow=k_\downarrow$ independent of
$\Delta N\equiv N_\uparrow-N_\downarrow$
\begin{eqnarray}\label{eqdualprobeapp}
    {\rm var}\left(\phi_\uparrow+\phi_\downarrow\right) &=&{\rm var}\left(k \Delta N\right) =\langle k\rangle^2\underbrace{{\rm var}\left(\Delta N\right)}_{N}+{\rm var}\left(k\right) \underbrace{\langle \Delta N\rangle^2}_{0}+{\rm var}\left(k\right) \underbrace{{\rm
  var}\left(\Delta N\right)}_{N}\nonumber\\&=&\left[\langle k\rangle^2+{\rm
  var}\left(k\right)\right]N,
\end{eqnarray}
using CSS properties ${\rm var}(N_\uparrow-N_\downarrow)=N$ and
$\langle N_\uparrow-N_\downarrow\rangle^2=0$ (cf. section
\ref{sec:atomprojnoise}).
\end{document}